\documentclass[letterpaper,twocolumn,10pt]{article}
\usepackage{usenix-2020-09}

\makeatletter
\@ifpackageloaded{hyperref}{%
  \hypersetup{colorlinks=true,allcolors=black}%
}{%
  \usepackage{hyperref}%
  \hypersetup{colorlinks=true,allcolors=black}%
}
\makeatother

\usepackage{tikz}
\usetikzlibrary{arrows.meta}
\usepackage{xspace}
\usepackage{balance}
\usepackage{threeparttable}
\usepackage{listings}
\usepackage[normalem]{ulem}
\usepackage{multirow}
\usepackage{svg}
\usepackage{pifont}
\usepackage{chronology}
\usepackage{timeline}
\usepackage{flushend}
\usepackage{booktabs}
\usepackage{algorithm,algorithmic}

\usepackage{tcolorbox}
\definecolor{mycolor}{rgb}{0.122, 0.435, 0.698}
\definecolor{gray1}{gray}{0.3}
\definecolor{darkgreen}{rgb}{0.0, 0.5, 0.0}
\definecolor{darkred}{rgb}{0.82, 0.1, 0.26}
\definecolor{shallowgreen}{RGB}{196, 214, 160}
\definecolor{shallowred}{RGB}{217, 149, 143}
\newcommand{\cmark}{\textcolor{darkgreen}{\ding{51}}}%
\newcommand{\result}[1]{%
\begin{tcolorbox}[colframe=mycolor,boxrule=0.5pt,arc=4pt,
      left=6pt,right=6pt,top=6pt,bottom=6pt,boxsep=0pt,width=\columnwidth]%
      {#1}
\end{tcolorbox}%
}

\newcommand{\sqlancer}{\textsl{SQLancer}\xspace}
\newcommand{\method}{\textsl{BFA}\xspace} 
\newcommand{\tool}{\textsl{QueryZen}\xspace}
\newcommand{\amoeba}{\textsl{AMOEBA}\xspace}
\newcommand{\apollo}{\textsl{APOLLO}\xspace}
\newcommand{\puppy}{\textsl{PUPPY}\xspace}
\newcommand{\cert}{\textsl{CERT}\xspace}
\newcommand{\todo}[1]{}
\renewcommand{\todo}[1]{{\color{red} TODO: {#1}}}
\newcommand{\add}[1]{}
\renewcommand{\add}[1]{{{#1}}}
\newcommand{\del}[1]{}
\renewcommand{\del}[1]{{\color{blue} \sout{#1}}}

\newcommand{\blind}[1]{#1}

\newcommand{\etal}{\textit{et al}.\xspace}
\newcommand{\ie}{\textit{i}.\textit{e}.\xspace}
\newcommand{\eg}{\textit{e}.\textit{g}.\xspace}

\newcommand{\numbugs}{21\xspace}
\newcommand{\numconfirmedbugs}{19\xspace}

\title{Finding Performance Issues in Database Systems\\ by Exploiting Dormant Code Paths}

\author{
Jinsheng Ba\\
ETH Zurich\\
Switzerland
\and
Zhendong Su\\
ETH Zurich\\
Switzerland
}

\begin{document}

\maketitle

\begin{abstract}
Performance is a critical characteristic of fundamental systems, such as Database Management Systems (DBMSs). Both academia and industry have invested decades in exploring efficient optimization algorithms. Despite these efforts, DBMSs are prone to performance issues, which incur suboptimal performance. Finding such issues is a longstanding challenge as no ground-truth performance is available. Existing work adopts \emph{black-box} methods to examine performance consistency across executions, but cannot systematically test optimizations. In this work, we propose a novel, general \emph{white-box} methodology, \emph{Branch Flip Analysis} (BFA), to systematically and effectively uncover performance issues. BFA flips code branches to enforce or disable an optimization, and the performance is expected to be not significantly better. Otherwise, a performance issue exists. BFA provides a new perspective to finding performance issues and testing optimization logics in a fine-grained manner. We realized BFA in a prototype system QueryZen, and evaluated it on four widely-used and mature DBMSs: PostgreSQL, MySQL, CockroachDB, and MariaDB. QueryZen found 21 previously unknown and unique performance issues with the workload of the extensively used benchmarks TPC-H and TPC-DS. The core concept of BFA is simple and broadly applicable, and can be adapted to analyze the performance of other software systems.
\end{abstract}

\section{Introduction}
Database Management Systems (DBMSs) serve as the backbone for data manipulation and processing across various applications. Ensuring optimal DBMS performance is critical for maintaining the efficiency, scalability, and reliability of the systems they support. Over the past decades, researchers and practitioners have explored various algorithms and architectures to optimize performance, including join order search space exploration~\cite{neumann2009query, fender2012effective, fender2013counter} and hardware acceleration~\cite{giceva2014deployment, paul2016gpl, wu2014q100}. However, \emph{performance issues} that incur suboptimal performance remain a persistent challenge.

Automatically finding \emph{performance issues} is challenging due to the complexity of modern DBMSs.  
First, determining whether executing a query on a database results in unexpectedly suboptimal performance is difficult, as there is typically no ground truth to define what constitutes optimal performance. This challenge is further compounded by the use of various heuristics and cost models during query optimization, which often involve trade-offs that prioritize optimizing certain types of queries at the expense of others.
Second, complex optimization strategies in DBMSs make it challenging to systematically test performance. For a given query, DBMSs may apply join optimization to determine the optimal join order, index optimization to select the appropriate indexes, and so on. These optimizations collectively influence performance. Relying solely on input queries, however, cannot comprehensively explore the full range of optimization strategies and their interactions.

Various black-box methods have been proposed to derive ``optimal'' performance for finding performance issues, but they focus on a specific category of issues and cannot systematically explore optimization strategies.
\apollo~\cite{jung2019apollo} examines whether a DBMS has worse performance than an old version of the same DBMS. However, \apollo can only find regression issues.
\amoeba~\cite{liu2022automatic} expects that the execution time of pairs of semantically equivalent queries is similar. However, it only finds the issues of rewriting queries~\cite{liu2022automatic}.
\cert~\cite{ba2024cert} finds performance issues by finding inconsistent cardinality estimation, which is typically deemed as the most critical component for query optimization~\cite{leis2015good}. However, \cert can only find performance issues related to incorrect cardinality estimation.
\puppy~\cite{wu2024puppy} assumes that the execution time of queries under the default optimization configuration should be no longer than under alternative configurations. However, this approach is limited to detecting configuration-related performance issues.
These methods focus on manipulating input queries, which are inefficient for exploring optimization strategies.

In this paper, we propose \emph{Branch Flip Analysis} (\method), a novel, general white-box method to systematically find performance issues by manipulating the DBMS source code. Although a DBMS’s default optimization strategy is not guaranteed to be optimal, it should still deliver near-optimal performance. Our idea is to selectively disable or enforce specific optimizations through code manipulation to construct a reference DBMS. This reference version should never significantly outperform the original. Otherwise, a performance issue exists in the original DBMS, as it should achieve at least comparable performance to the reference DBMS. We propose a simple and general code manipulation operation that flips code branches of \emph{IF} statements, which we observed are commonly used for optimization decisions.
Formally, we define \method as follows. For a DBMS's source code, $B = \{b_1,b_2,...,b_n\}$ is the set of all conditional branches of the \emph{IF} statements that enable or disable optimization strategies. For a branch $b_i \in B$, let $b_{i,d}$ and $b_{i,o}$ denote, respectively, the default and the alternative execution path. Given a specific workload, $P(B)$ represents the performance of the DBMS with $B$. For a specific branch $b_j \in B$, we flip its branch and derive $B^\prime$ such that $\forall i \neq j\ (bi=b_{i,d}) \& b_j = b_{j,o}$, \ie, all branches except $b_j$ execute their default paths $b_{i,d}$, while $b_j$ executes the alternative path $b_{j,o}$. If $P(B) << P(B^\prime)$, a performance issue related to the optimization controlled by $b_j$ is found.

Thus, in a novel manner, \method solves the aforementioned challenges faced by existing work---it derives expected performance by constructing a reference DBMS. Code manipulation also allows us to explore various optimization strategies in a systematic and fine-grained manner. In contrast, existing work is constrained by external inputs or configurations, which often fail to control all the available optimizations. To manipulate code, we propose flipping branches, which is motivated by our observation that 8 of 10 historical performance issues from \apollo, \amoeba, \cert, and \puppy could be found by flipping branches. Branch flipping is straightforward to realize and general, so it can be widely applied to test DBMSs without domain expertise. Additionally, it highlights the code segment relevant to the performance issue, thus facilitating issue analysis. 

\begin{figure}
\begin{lstlisting}[caption={A performance issue found by \method in PostgreSQL.},captionpos=t, label=lst:motivation, escapeinside=&&, language=sql]
SELECT ... FROM PART, SUPPLIER, PARTSUPP ...
WHERE p_partkey = ps_partkey AND s_suppkey = ps_suppkey ...;
------------------Code Manipulation------------------
--- a/src/backend/optimizer/util/clauses.c
+++ b/src/backend/optimizer/util/clauses.c
@@ -803,8 +803,6 @@ max_parallel_hazard_test(...)
  Assert(context->max_hazard != PROPARALLEL_UNSAFE);
  context->max_hazard = proparallel;
  /* done if we are not expecting any unsafe functions */
- if (context->max_interesting == proparallel)
-   return true;
  break;
-------------Query Plans and Performance-------------
...                    ...
->HashJoin(cost=66839) ->Parallel HashJoin(cost=41019)
  ->Parallel Scan part   ->Parallel Scan part
  ->Hash                 ->Parallel Hash
    ->HashJoin             ->HashJoin 
...                    ...
Time: 266ms            Time: 130ms
\end{lstlisting}
\end{figure}

\autoref{lst:motivation} shows a running example of \method. Lines 1--3 include the query to identify the performance issue, lines 4--12 show the code that \method manipulates, and lines 14--19 show the query plans, the concrete execution steps of this query in a tree structure, before and after the code manipulation. To facilitate the presentation, we only show relevant parts. This query includes three-table joining. \lstinline{SUPPLIER} is joined with \lstinline{PARTSUPP} by \lstinline{HashJoin}, as shown in the left part of line 18. Their result is joined with \lstinline{PART} by \lstinline{HashJoin} again, as shown in the left part of line 15. The \emph{IF} statement in line 10 checks whether to ignore \lstinline{PARALLEL} optimization, which is an optimization strategy in PostgreSQL that executes an operation in multiple CPU cores for efficiency. For this query, this optimization only applies to the scanning of \lstinline{PART} in line 16. We flipped the \emph{IF} branch, so the \lstinline{PARALLEL} optimization is applied to other operations \lstinline{HashJoin} and \lstinline{Hash}. We found that PostgreSQL has better performance without breaking functionality. Note that the code patch used by \method is an overfix as it enforces \lstinline{PARALLEL} for all operations. To properly fix this issue, we require developers' expertise to determine which operations use \lstinline{PARALLEL}. When we reported this issue to developers, they replied that ``\emph{It was a pretty exciting case}'' and planned to solve it by enabling \lstinline{PARALLEL} for these operations shortly.  This issue can not be found by previous methods because \lstinline{PARALLEL} cannot be enabled for \lstinline{Hash} by any version of PostgreSQL (\ie, regression issues), any configuration, or any estimated cardinality.

We designed and implemented \tool, a prototype of \method. Technically, to avoid the heavy overhead of recompilation for each code manipulation, we instrument all code manipulations into the code with a single compilation and enable them one by one via external environmental variables. 

We evaluated \tool on PostgreSQL, MySQL, MariaDB, and CockroachDB, which are widely used in previous performance testing works~\cite{jung2019apollo, wu2024puppy, ba2024cert}. In total, \tool found \numbugs unique and previously unknown performance issues with the workloads from standard benchmarks (\ie, TPC-H~\cite{tpch} and TPC-DS~\cite{tpcds}). These findings resulted in an average performance improvement of 26.2$\times$, and up to 374.9$\times$. \tool is necessary to find these issues as most of them cannot be found by \apollo, \cert, \puppy, and \amoeba. Furthermore, these issues already existed before the publication of prior methods. For testing efficiency, we avoid the recompilation time of an average of 5 minutes for each execution. On average of both benchmarks, \tool also excludes 75.8\% unnecessary code manipulations and increases branch coverage by 8.6\%.

Overall, we make the following contributions:
\begin{itemize}
    \item Conceptually, we propose a novel methodology \method to find performance issues by flipping code branches.
    \item Technically, we designed and implemented a novel architecture \tool to avoid recompilation and unnecessary code manipulations for efficient testing.
    \item Practically, we evaluated \tool on four widely-used DBMSs, and found \numbugs previously unknown and unique performance issues. 
\end{itemize}

\section{Background}

\paragraph{Query optimization.}
Given a query $Q$ in a declarative language, a DBMS parses and translates it into a concrete execution plan $P$, also known as a query plan. To optimize $P$, DBMSs typically use heuristic rules to transform $Q$ to an equivalent $Q^\prime$, expecting that the execution cost of $P^\prime$, the plan of $Q^\prime$, is lower than that of $P$~\cite{jarke1984query, ioannidis1996query, chaudhuri1998overview}. For instance, if $Q$ contains a projection operator $\pi$ that selects specific columns $A$, pushing $\pi$ down can eliminate unnecessary columns early: $Q:\pi_A(R \bowtie S) \Rightarrow Q^\prime: \pi_A(\pi_A(R) \bowtie \pi_A(S))$, where $\bowtie$ represents joining tables $R$ and $S$. However, this rule may introduce potential performance issues. If $Q^\prime$ includes the operators (\eg, \lstinline{GROUP BY}) that require the columns removed by $\pi$, recomputation may be necessary incurring degraded performance. 
Additionally, modern DBMSs employ cost-based optimization, deriving $P$ to $n$ semantically equivalent query plans $\{P_1,P_2,...,P_n\}$, and estimating the execution cost of each $P_i$ by the cost function $C(P_i)$~\cite{leis2015good}. DBMSs choose and execute the optimal plan $P_o = \arg\min_{P_i \in P} C(P_i)$. Incorrect $C()$ brings potential performance issues.

\section{Motivating Study}\label{sec:study}
To investigate the potential of finding performance issues by code manipulation, we studied whether we can manipulate code to observe the unexpected performance of historical issues found by prior methods. 

\begin{table}
    \centering\small
    \caption{Historical performance issues.}
    \begin{tabular}{@{}llrc@{}}
    \toprule
    \textbf{Method} & \textbf{DBMS} & \textbf{Issue ID} & \textbf{Branch Flipping} \\
    \midrule
    \apollo  & PostgreSQL   & \href{https://www.postgresql.org/message-id/BN6PR07MB3409EE6CAAF8CCF43820AFB9EE670%40BN6PR07MB3409.namprd07.prod.outlook.com}{1}         & \cmark \\  
    \apollo  & PostgreSQL   & \href{https://www.postgresql.org/message-id/BN6PR07MB3409EE6CAAF8CCF43820AFB9EE670%40BN6PR07MB3409.namprd07.prod.outlook.com}{3}         & \cmark \\  
    \amoeba  & CockroachDB  & \href{https://github.com/cockroachdb/cockroach/issues/51820}{51820}   & \cmark \\  
    \amoeba  & CockroachDB  & \href{https://github.com/cockroachdb/cockroach/issues/57330}{57330}   &            \\  
    \amoeba  & CockroachDB  & \href{https://github.com/cockroachdb/cockroach/issues/57566}{57566}   & \cmark \\  
    \amoeba  & CockroachDB  & \href{https://github.com/cockroachdb/cockroach/issues/58283}{58283}   & \cmark \\  
    \amoeba  & CockroachDB  & \href{https://github.com/cockroachdb/cockroach/issues/58284}{58284}   &            \\  
    \cert    & CockroachDB  & \href{https://github.com/cockroachdb/cockroach/issues/88455}{88455}   & \cmark \\  
    \cert    & CockroachDB  & \href{https://github.com/cockroachdb/cockroach/issues/89161}{89161}   & \cmark \\  
    \puppy   & MySQL        & \href{https://bugs.mysql.com/bug.php?id=115676}{115676}               & \cmark \\  
    \bottomrule
     & & & \textbf{Sum: 8/10} \\
    \end{tabular}
    \label{tab:historical_issues}
\end{table}

\paragraph{Issue collection.}
We considered the performance issues found by the four existing performance-testing methods: \apollo~\cite{jung2019apollo}, \amoeba~\cite{liu2022automatic}, \cert~\cite{ba2024cert}, and \puppy~\cite{wu2024puppy}. \apollo examines performance regression across different versions of DBMSs; \amoeba expects semantic-equivalent queries to have similar performance; \cert checks specifically incorrect estimated cardinalities, which refers to the estimated number of rows returned by a query plan; \puppy compares performance under the default configuration and other configurations.
In total, we identified and reproduced 10 confirmed or fixed performance issues. 
\apollo has 3 confirmed performance issues. Although it does not publicize the full list of found issues, we found these issues in DBMS bug trackers. However, due to the environment (\eg, database download link) in the issue reports being invalid now, we successfully reproduced 2 issues in our environment. 
\amoeba has 6 confirmed issues. Similar to \apollo, we reproduced 5 issues.
\cert has 11 confirmed issues. Unlike other methods, \cert specifically detects inconsistent estimated cardinalities, instead of execution time. In Listing 7 of \cert paper, authors provide two cases to show the execution time gap between pairs of queries due to incorrect cardinalities found, and we evaluated both cases.
\puppy has 54 confirmed issues, but does not publicize the full list of its found issues. By searching the bug trackers of DBMSs, we found a report, which we believe is similar to the issues described in \puppy paper.

\paragraph{Results.}
\autoref{tab:historical_issues} shows the 10 performance issues from \apollo, \cert, \puppy, and \amoeba. Overall, we can observe the unexpected performance of 8 issues by the same code manipulation that flips code branches of \emph{IF} statements once. Specifically, we changed \lstinline{if (condition)} to \lstinline{if (!condition)}. 
Both performance issues in \apollo are due to the redundant initialization of hash tables and incorrect parallel computing. Flipping branches disables both, so we can observe the unexpected performance. 
Both performance issues in \cert affect the estimated cardinalities incurring an inefficient usage of \lstinline{HASH JOIN}. Flipping branches changes the strategies of cardinality estimation to find both issues.
\puppy affects optimization strategies by configurations, whose values are checked in \emph{IF} statements, so flipping branches can achieve the same effect as configurations.
For \amoeba, we cannot use code manipulation to find \#57330 and \#58284, because both performance issues are due to unimplemented optimizations, such as simplifying \lstinline{JOIN ON TRUE} to \lstinline{JOIN}.
Overall, this simple code manipulation of flipping branches has proven effective in identifying historical performance issues found by various testing methods, demonstrating its potential for finding performance issues through code manipulation.

\begin{figure}
\begin{lstlisting}[caption={Historical performance issue \href{https://bugs.mysql.com/bug.php?id=115676}{115676} in MySQL.},captionpos=t, label=lst:study, escapeinside=@@]
SET optimizer_switch = 'batched_key_access=on';
CREATE TABLE t0(c1 DECIMAL ZEROFILL, c2 FLOAT UNIQUE);
INSERT DELAYED IGNORE INTO t0(c2, c1) VALUES...;
CREATE TABLE t2(c1 FLOAT) ;
INSERT LOW_PRIORITY IGNORE INTO t2(c1) VALUES("6Y^c!3Sq")...;
CREATE TABLE t3(c0 TINYINT UNIQUE KEY, c2 INT(252));
CREATE TABLE IF NOT EXISTS t4(c0 DECIMAL, c2 FLOAT, c10 DECIMAL) ;
CREATE TABLE t5 LIKE t4;
INSERT LOW_PRIORITY IGNORE INTO t4(c0) VALUES(NULL)...;

SELECT DISTINCT t0.c1 FROM t4, t2, t0, t3 NATURAL JOIN t5; -- 28ms
SELECT /*+ BKA()*/ DISTINCT t0.c1 FROM t4, t2, t0, t3 NATURAL JOIN t5; -- 22ms
------------------Code Manipulation------------------
--- a/sql/sql_optimizer.cc
+++ b/sql/sql_optimizer.cc
@@ -3511,7 +3511,7 @@ static bool setup_join_buffering(JOIN_TAB *tab, JOIN *join,
     return false;
   }
-  if (!(bnl_on || bka_on)) goto no_join_cache;
+  if ((bnl_on || bka_on)) goto no_join_cache;
 
\end{lstlisting}
\end{figure}

\paragraph{Example.}
\autoref{lst:study} shows the issue \#115676 in MySQL found by \puppy. Lines 1--9 include the SQL statements to create the database, lines 11--12 show the queries to identify the performance issue, and lines 14--20 show MySQL source code. The issue is due to improper usage of Batched Key Access (BKA) optimization, which decides whether to use indexes for joining. The second query in line 12 uses the query hint \lstinline{/*+ BKA()*/} to enforce BKA incurring a shorter execution time than the first query in line 11, which \puppy finds as a performance issue. In line 19, the \emph{IF} statement checks the variable \lstinline{bnl_on}, which stores the value of the configuration BKA. In default, the value is false, so \lstinline{goto no_join_cache} is not executed. We flipped this branch to execute \lstinline{goto no_join_cache}, and it is equivalent to enabling BKA. Therefore, manipulating code can observe the same unexpected performance found by changing configurations.

\result{8 of 10 historical performance issues can be found by flipping the code branch of an \emph{IF} statement.}


\section{Approach}
We propose \method, a novel approach to identify performance issues by manipulating program code. The core intuition behind \method is that the default optimization strategy should ideally have optimal performance. If removing or enforcing an optimization brings in a performance improvement, it indicates a potential performance issue in the target DBMS.

\begin{figure*}
    \centering
    \includegraphics[width=\textwidth]{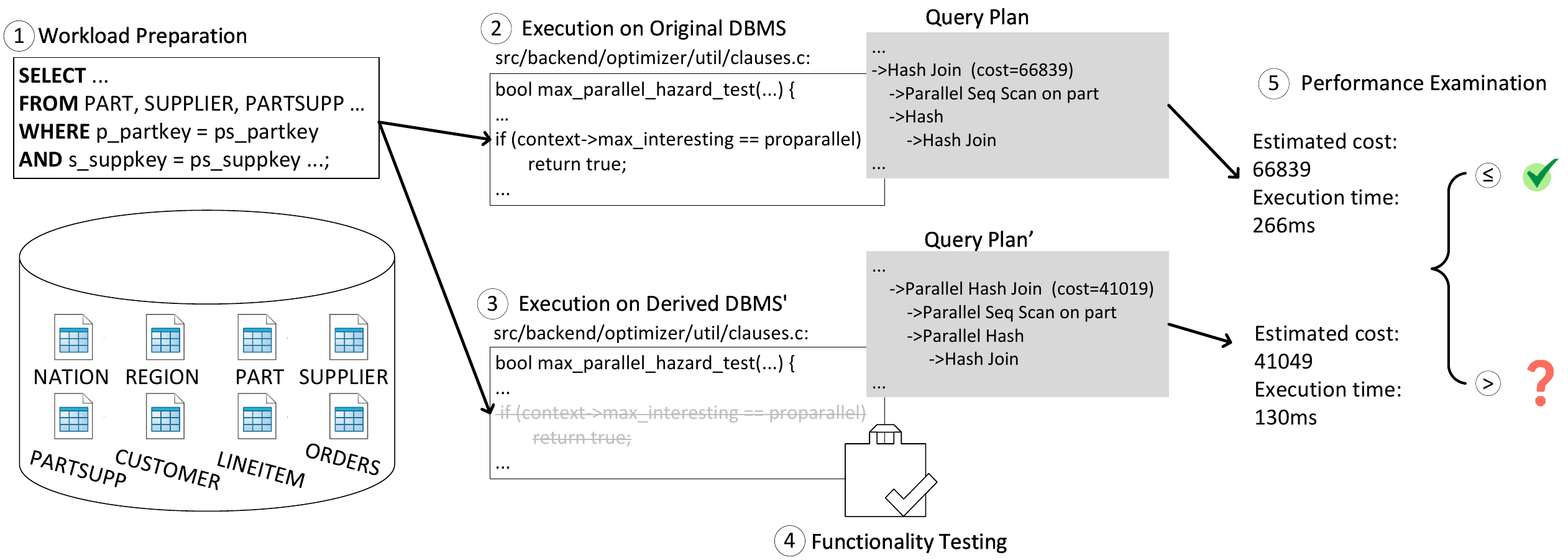}
    \caption{Overview of \method.}
    \label{fig:approach}
\end{figure*}

\paragraph{System overview.}
\autoref{fig:approach} provides an overview of \method, illustrated using the running example in \autoref{lst:motivation}. The process begins in step \textcircled{1}, where we prepare a database and a query to serve as the workload. In step \textcircled{2}, we execute the target DBMS on the workload. Next, in step \textcircled{3}, we manipulate the program code to derive an alternative version, referred to as DBMS$^\prime$, and execute DBMS$^\prime$ on the same workload. 
To ensure that the manipulated code in DBMS$^\prime$ only affect performance optimization, step \textcircled{4} validates whether DBMS and DBMS$^\prime$ produce semantically consistent results. Finally, step \textcircled{5} evaluates the performance of DBMS and DBMS$^\prime$. If DBMS$^\prime$ demonstrates superior performance, it shows a potential performance issue in the original DBMS.

\subsection{Workload Preparation (\textcircled{1})}
In principle, \method can test target DBMSs with any workload. In DBMSs, a \emph{workload} refers to the collection of tasks or operations that DBMSs need to execute over a given period. In this paper, we use \emph{workload} to represent databases and queries specifically. In \autoref{fig:approach}, we used the TPC-H benchmark~\cite{tpch} to populate the database with multiple tables including \lstinline{PART, SUPPLIER,} and \lstinline{PARTSUPP}, and the query is provided by the benchmark. To facilitate the presentation, we simplified the query in our example to highlight only the clauses relevant to the performance issue.

\subsection{Execution on Original DBMS (\textcircled{2})}
We execute the target DBMS with the workload obtained in step \textcircled{1}. Before executing the query, the DBMS translates it into a query plan, which is then executed to produce results. During this process, we record both the estimated cost from the query plan and the execution time, which are later utilized in step \textcircled{5}. In \autoref{fig:approach}, the query involves joining the tables \lstinline{PART}, \lstinline{SUPPLIER}, and \lstinline{PARTSUPP}. The query plan employs the \lstinline{HASH JOIN} operation to perform joining and \lstinline{Seq Scan} operation to read data from tables. PostgreSQL, as the target DBMS, leverages an optimization strategy that utilizes parallel CPU cores to enhance performance. Specifically, for this query, PostgreSQL executes the table scan of \lstinline{PART} in parallel using the \lstinline{Parallel Seq Scan on part} operation.

\subsection{Execution on Derived DBMS$^\prime$ (\textcircled{3})}
In this step, we manipulate DBMS's source code to derive DBMS$^\prime$ and execute DBMS$^\prime$ on the same workload as DBMS.

Considering testing efficiency, our code manipulations focus on the query optimization component. DBMSs, such as PostgreSQL, are highly complex systems; for instance, PostgreSQL version 17 consists of over 2 million lines of code. Among these, query optimization is a critical area directly tied to performance improvement. As shown in \autoref{fig:approach}, the query optimization code in PostgreSQL is located within the folder \emph{src/backend/optimizer}, which we specifically targeted for our manipulations. Note that focusing on a particular component is not compulsory but for efficiency.

Followed by our study in \autoref{sec:study}, we propose a simple strategy to manipulate code by flipping the branches of \emph{IF} statements. For both branches of an \emph{IF} statement, if DBMS executes one, we request DBMS$^\prime$ to execute the other one. Suppose DBMS applies an optimization in the executed branch of an \emph{IF} statement, DBMS$^\prime$ disables this optimization by executing the other branch, and vice versa. We only manipulate an \emph{IF} statement per execution. We believe this strategy is simple yet efficient for exploring optimization space in code. In \autoref{fig:approach}, DBMS executes the true branch of \lstinline{if (context->max_interesting == proparallel)}, so we disable the true branch and enforce DBMS$^\prime$ to execute the false branch.

\subsection{Functionality Testing (\textcircled{4})}
To make sure the flipped branches only enable or disable optimization strategies with no side effects, we check whether DBMS$^\prime$ is still functionally correct. We propose two solutions for DBMS users and developers, respectively.

As DBMS users lack expert knowledge of the DBMS$^\prime$s internal code, we adopted a differential testing approach to evaluate functionality in practice. Differential testing involves comparing the results of executing the same inputs across different systems; inconsistencies in the results may indicate potential bugs. Specifically, we validate whether DBMS and DBMS$^\prime$ consistently produce identical results. If they do, we assume that DBMS$^\prime$ is functionally correct. The trustworthiness of this method increases with the number of test cases evaluated. In this paper, we collected 10 thousand randomly generated small workloads from \sqlancer~\cite{Rigger2020TLP, Rigger2020NoREC}, which is a popular DBMS testing tool. As shown in our evaluation (\autoref{sec:evaluation}), 96\% of the DBMS$^\prime$s that pass the above validation do not affect program functionalities.

For DBMS developers, we can use specification-based validation approaches. Since developers have expert knowledge of the code, we expect developers to provide a specification of the query plans, which can be used for checking validity. For example, in PostgreSQL's query plans, the operation \lstinline{GroupAggregate} aggregates pre-sorted rows for a GROUP BY operation, so its child operation must be \lstinline{Sort}. Such a rule can be used to automatically check whether the query plans produced by DBMS$^\prime$ are correct. 

\subsection{Performance Examination (\textcircled{5})}
Lastly, for the DBMS$^\prime$ that are functionally correct, we compare the performance of DBMS and DBMS$^\prime$ obtained from steps \textcircled{2} and \textcircled{3}. In addition to execution time, we also compare the estimated cost for testing efficiency. 
Measuring execution time may take several minutes per query. For high testing efficiency, before executing a query to measure its execution time in steps \textcircled{2} and \textcircled{3}, we obtain the query plan and compare the estimated cost, which can be obtained by executing the query with the prefix \lstinline{EXPLAIN} in milliseconds. Only when DBMS$^\prime$ has a smaller estimated cost than DBMS, do we execute the query to evaluate the execution time. 
Measuring estimated cost significantly improves the testing efficiency but may miss the cases that DBMS$^\prime$ reports a higher estimated cost but a shorter execution time compared to DBMS. Therefore, we leave it optional for \method. When finding a potential performance issue, our method outputs the workload, a code patch equivalent to the code manipulation, both query plans, and execution time. In \autoref{fig:approach}, DBMS$^\prime$ has a significantly smaller estimated cost and execution time than DBMS, so it indicates a potential performance issue, which was quickly confirmed by PostgreSQL developers.
\section{Implementation}
We implemented \method in the \tool prototype. We discuss the key implementation details in this section.

\subsection{Workload Selection for \textcircled{1}}
We used the standard benchmarks TPC-H~\cite{tpch} and TPC-DS~\cite{tpcds} as workloads. While exploring diverse workloads can reveal more performance issues, it is more practical to focus on commonly used workloads. Unlike crashes, where any issue in any workload can have significant consequences, performance issues are more impactful when they affect widely used workloads. Developers generally consider an optimization beneficial if it improves performance on commonly used workloads, even if it causes regressions on rarely used ones~\cite{postgresqlworkload}. For this reason, we adopted standard benchmarks TPC-H and TPC-DS. TPC-H is a decision support benchmark simulating a wholesale supplier’s sales scenario, featuring 22 queries, while TPC-DS is a more advanced benchmark representing a large-scale retail business scenario, comprising 99 queries. Both benchmarks are widely used for performance evaluation~\cite{ivanov2017adaptive, boncz2013tpc}, and we believe their workloads are representative of commonly used scenarios.

\subsection{Code Manipulation for \textcircled{3}}
To efficiently manipulate code, which is the critical step for \method, we did two optimizations to eliminate recompilation and unnecessary code manipulations.

\paragraph{Eliminating recompilation.}
For each code manipulation, the recompilation of the target DBMS often demands significant computational resources, especially for complex DBMSs. This poses a substantial limitation to the efficiency of \method. To address this challenge, we instrumented code manipulation operations into the program upfront and enabled each one dynamically using external environment variables. Specifically, for an \emph{IF} statement \lstinline{if (condition)}, we modified it to \lstinline{if (condition ^ get_env("DROP") == ID)}. Here, the function \lstinline{get_env("DROP")} retrieves the value of the environment variable \lstinline{DROP}, and \lstinline{ID} is a unique identifier for each \emph{IF} statement. The ``\lstinline{^}'' symbol denotes the bitwise XOR operator, which evaluates to true when both operands are not equal. If the value of \lstinline{DROP} matches \lstinline{ID}, the \emph{IF} condition is negated, causing the other branch to execute. For other values of \lstinline{DROP}, the \emph{IF} statement remains unaffected. Consequently, we can control the behavior of each \emph{IF} statement by setting the environment variable \lstinline{"DROP"} appropriately, eliminating the need for recompilation during testing. Although it introduces additional performance overhead, the overhead is tiny. For example, our instrumentation introduces less than 0.1 second additional execution time for PostgreSQL when running the TPC-H benchmark. More importantly, the performance overhead is added to both DBMS and DBMS$^\prime$, so it does not affect performance comparison.

\paragraph{Eliminating unnecessary code manipulation.}
When executing a query, many \emph{IF} statements are not executed, and manipulating them is inefficient. It is especially wasteful for large-scale DBMS source code, which includes thousands \emph{IF} statements to flip. To further enhance testing efficiency, we record the \emph{IF} statements covered during execution and manipulate only those that are executed. To achieve this, we add a logging function to the instrumented \emph{IF} statement as follows: \lstinline{if((condition) ^ (log(ID) && get_env("DROP") == ID))}. The function \lstinline{log(ID)} logs the \lstinline{ID} of the statement to a log file, allowing us to track which \emph{IF} statements were executed. This information helps us selectively decide which \emph{IF} statements to manipulate. Importantly, the \lstinline{log(ID)} function always returns true, ensuring it does not alter the semantics of the instrumented \emph{IF} statement.

\renewcommand{\algorithmicrequire}{\textbf{Input:}}
\renewcommand{\algorithmicensure}{\textbf{Output:}}
\begin{algorithm}[h]
\caption{\method implementation in \tool}
\label{alg:recondition}
\begin{algorithmic}[1]
\REQUIRE {program: $DBMS$, program: $DBMS^\prime$, query: $Q$}
\STATE{$cost = explain(DBMS, Q)$}
\STATE{$time, if\_list = execute(DBMS, Q)$}
\FOR{$ID$ in $if\_list$}
\STATE{SET DROP = ID}
\STATE{$cost^\prime = explain(DBMS^\prime, Q)$}
\IF{$cost > cost^\prime$}
\STATE{$time = execute(DBMS^\prime, Q)$}
\STATE{$compare(time, time^\prime)$}
\ENDIF
\ENDFOR
\end{algorithmic}
\end{algorithm}

\autoref{alg:recondition} shows the implementation algorithm of \method in \tool that eliminates recompilation and unnecessary code manipulations. First, we obtain the estimated cost of $Q$ on DBMS in line 1, and execute $Q$ to obtain execution time and the list of executed \emph{IF} statements in line 2. Then, we enumerate the list and enable each corresponding code manipulation by setting the environmental variable \lstinline{DROP} in lines 3 and 4. After that, if the estimated cost of executing $Q$ on DBMS$^\prime$ is smaller than that on DBMS, we execute $Q$ on DBMS$^\prime$ to check whether $Q$ has a shorter execution time. If so, a potential performance issue is found. This algorithm ensures that we only manipulate executed code without recompilation, significantly improving testing efficiency.


\section{Evaluation}\label{sec:evaluation}

To evaluate the effectiveness of \tool in finding performance issues, we sought to answer the following questions:
\begin{description}
    \item[\textbf{Q.1 Effectiveness.}] Can \tool find previously unknown performance issues?
    \item[\textbf{Q.2 Necessity.}] Whether \tool is necessary to find these performance issues?
    \item[\textbf{Q.3 Sensitivity.}] To what extent \tool's components contribute to the finding of these performance issues?
\end{description}

\paragraph{Tested DBMSs.}
We tested PostgreSQL, MySQL, CockroachDB, and MariaDB. PostgreSQL is one of the most established DBMSs. MySQL is the most popular open-source DBMS~\cite{dbranking}. MariaDB is another popular DBMS that was forked from MySQL. CockroachDB is a widely used enterprise-class DBMS, and its open version on GitHub has been starred more than 30.1k times. These DBMSs have been extensively tested by previous works about finding performance issues~\cite{ba2024cert, wu2024puppy, liang2024mozi, jung2019apollo, liu2022automatic}. Recall that \tool is not specific to any programming language, so we can test both C (PostgreSQL, MySQL, MaraiDB) and Go (CockroachDB) programming languages. We used the latest available development versions of tested DBMSs (PostgreSQL: 17.0, MySQL: 9.0, MariaDB: 11.7, CockroachDB: dcb0d27). We manipulated the code in specific folders (PostgreSQL: \emph{/app/postgres/src/backend/optimizer}, MySQL: \emph{/app/mysql-server/sql}, CockroachDB: \emph{/app/cockroach/pkg/sql/opt}, MariaDB: \emph{/app/mariadb-server/sql}). The folders of PostgreSQL and CockroachDB include only the code of query optimizer, and the folders of MySQL and MariaDB include all SQL-related code.

\paragraph{Benchmark preprocess.}
We adapted TPC-H and TPC-DS benchmarks for the four tested DBMSs and removed incompatible queries. TPC-DS benchmark includes 99 test cases. Each test case includes one or multiple queries, and we only use the first query in the evaluation for simplicity. Additionally, due to incompatible SQL grammar, we removed 3, 7, 14, and 12 cases in the TPC-DS benchmark for PostgreSQL, MySQL, CockroachDB, and MariaDB, respectively. TPC-H includes 22 test cases, which can be directly executed in the four tested DBMSs. We excluded case 15 because it includes creating views, while \tool aims to evaluate queries. We used 1 GB of data for both benchmarks, considering it to be a balanced trade-off between execution time and testing efficiency. This choice aligns with prior work~\cite{ivanov2017adaptive}, which employed the same data size.

\paragraph{Experimental infrastructure.}
We conducted all experiments on an AMD Ryzen Threadripper 3990X processor that has 64 physical and 128 logical cores clocked at 2.2 GHz. Our test machine uses Ubuntu 24.04 with 256 GB of RAM. We limit our maximum utilization to 20 physical cores to avoid resource competition affecting performance measurement.

\subsection*{Q.1 Effectiveness}

\begin{table}
    \centering\footnotesize
    \caption{Previously unknown and unique performance issues found by \tool. }
    \begin{tabular}{@{}llrrrl@{}}
        \toprule
        \textbf{DBMS} & \textbf{Query} & \textbf{Issue ID} & \textbf{Perf. $\uparrow$} & \textbf{Status} &\textbf{Num} \\
        \midrule
PostgreSQL  & TPC-H 2   & \blind{\href{https://www.postgresql.org/message-id/SEZPR06MB6494BD3DDF5B03700032C2C98A782\%40SEZPR06MB6494.apcprd06.prod.outlook.com}{1}  } & 2.0$\times$   & Confirmed & 2  \\ 
PostgreSQL  & TPC-H 2   & \blind{\href{https://www.postgresql.org/message-id/SEZPR06MB649480C104C2479D0F0D8ACD8A512\%40SEZPR06MB6494.apcprd06.prod.outlook.com}{2}  } & 3.1$\times$   & Confirmed & 6  \\ 
PostgreSQL  & TPC-H 10  & \blind{\href{https://www.postgresql.org/message-id/SEZPR06MB64943EFCD432E8069D610E458A4E2\%40SEZPR06MB6494.apcprd06.prod.outlook.com}{3}  } & 1.5$\times$   & Confirmed & 8  \\ 
PostgreSQL  & TPC-DS 4  & \blind{\href{https://www.postgresql.org/message-id/SEZPR06MB6494F6A2837995BDD4E0BF9A8A5F2\%40SEZPR06MB6494.apcprd06.prod.outlook.com}{4}  } & 374.9$\times$ & Confirmed & 11 \\ 
PostgreSQL  & TPC-DS 60 & \blind{\href{https://www.postgresql.org/message-id/SEZPR06MB649422CDEBEBBA3915154EE58A232\%40SEZPR06MB6494.apcprd06.prod.outlook.com}{5}  } & 1.5$\times$   & Fixed     & 8  \\ 
MySQL       & TPC-H 8   & \blind{\href{https://bugs.mysql.com/bug.php?id=116456}{116456}                                                                           } & 1.1$\times$   & Confirmed & 9  \\ 
MySQL       & TPC-H 11  & \blind{\href{https://bugs.mysql.com/bug.php?id=116484}{116484}                                                                           } & 3.1$\times$   & Confirmed & 9  \\ 
MySQL       & TPC-H 16  & \blind{\href{https://bugs.mysql.com/bug.php?id=116309}{116309}                                                                           } & 3.5$\times$   & Confirmed & 8  \\ 
MySQL       & TPC-H 16  & \blind{\href{https://bugs.mysql.com/bug.php?id=116534}{116534}                                                                           } & 2.1$\times$   & Confirmed & 7  \\ 
MySQL       & TPC-DS 7  & \blind{\href{https://bugs.mysql.com/bug.php?id=116773}{116773}                                                                           } & 12.1$\times$  & Confirmed & 12 \\ 
MySQL       & TPC-DS 15 & \blind{\href{https://bugs.mysql.com/bug.php?id=116774}{116774}                                                                           } & 2.4$\times$   & Confirmed & 24 \\ 
MySQL       & TPC-DS 16 & \blind{\href{https://bugs.mysql.com/bug.php?id=116775}{116775}                                                                           } & 1.7$\times$   & Confirmed & 22 \\ 
MySQL       & TPC-DS 19 & \blind{\href{https://bugs.mysql.com/bug.php?id=116776}{116776}                                                                           } & 1.9$\times$   & Confirmed & 9  \\ 
MySQL       & TPC-DS 47 & \blind{\href{https://bugs.mysql.com/bug.php?id=116777}{116777}                                                                           } & 1.6$\times$   & Confirmed & 9  \\ 
CockroachDB & TPC-H 2   & \blind{\href{https://github.com/cockroachdb/cockroach/issues/134803}{134803}                                                             } & 24.0$\times$  & Confirmed & 6  \\ 
CockroachDB & TPC-H 7   & \blind{\href{https://github.com/cockroachdb/cockroach/issues/135001}{135001}                                                             } & 2.9$\times$   & Confirmed & 5  \\ 
CockroachDB & TPC-DS 63 & \blind{\href{https://github.com/cockroachdb/cockroach/issues/136350}{136350}                                                             } & 2.5$\times$   & Confirmed & 12 \\ 
MariaDB     & TPC-H 2   & \blind{\href{https://jira.mariadb.org/browse/MDEV-35280}{35280}                                                                          } & 102.2$\times$ & Confirmed & 23 \\ 
MariaDB     & TPC-H 4   & \blind{\href{https://jira.mariadb.org/browse/MDEV-35332}{35332}                                                                          } & 2.6$\times$   & Analyzing & 18 \\ 
MariaDB     & TPC-H 7   & \blind{\href{https://jira.mariadb.org/browse/MDEV-35331}{35331}                                                                          } & 1.6$\times$   & Confirmed & 11 \\ 
MariaDB     & TPC-H 18  & \blind{\href{https://jira.mariadb.org/browse/MDEV-35333}{35333}   } & 1.9$\times$   & Analyzing & 4  \\                                                                                                                                          
        \bottomrule
        & & & & \textbf{Sum: \numbugs}
    \end{tabular}
    \label{tab:issues}
\end{table}

We ran \tool to find performance issues. For each potential performance issue, to exclude false alarms due to environmental factors, we recompiled the DBMS with the manipulated code in a separate environment and ran the same query ten times to check whether we could observe the same performance gap. After that, we reported our workload and code patch to the developers. \add{On average, each query requires 10 minutes to validate and deduplicate performance issues before reporting to developers.} 

\paragraph{Finding overview.} \autoref{tab:issues} lists the performance issues \tool found. \textbf{Perf. $\uparrow$} represents the performance gap due to the code manipulation. Developers typically respond directly regarding the status of the issue: \emph{Confirmed} denotes that the issue has been identified as a performance issue; \emph{Fixed} denotes that a fix has been proposed or pushed; \emph{Analyzing} denotes that developers are still investigating and have not yet replied to us. \textbf{Num} indicates the number of other TPC-H and TPC-DS queries affected by the same issue, with at least a 10\% performance gap. In total, \tool found \numbugs unique and previously unknown performance issues. Within them, \numconfirmedbugs issues had been confirmed, and one issue had been fixed by developers. These confirmed issues support our assumption that flipping branches effectively disables or enables optimization strategies to disclose performance issues. The remaining two issues in MariaDB were still under analysis, which may be due to the difficulty of analyzing performance issues. To avoid burdening developers' workload, we did not continue to report more issues in MariaDB. \tool found only three issues in CockroachDB, and a possible reason is that many optimizations are implemented in its domain-specific language, which \tool does not instrument. On average, these performance issues show a 26.2$\times$ performance gap compared to executing original queries. Especially, issue \blind{\#4} in PostgreSQL shows a 374.9$\times$ performance gap. Furthermore, these performance issues have a broader impact as these issues affect 10 other queries on average.

\paragraph{Root-reason analysis.}
We analyzed the root causes of these performance issues. We focused on PostgreSQL, whose developers were the most responsive in acknowledging and analyzing our reported issues. For the other three systems, detailed analysis was not available, likely due to their internal development policies or cycles. Within five confirmed issues in PostgreSQL, three are due to an underperforming executor, and the others are due to an inaccurate cost model. The executor-related issues stem from suboptimal implementation of certain execution strategies, while the cost model issues arise from inaccurate cost estimations that lead the optimizer to select suboptimal query plans. \autoref{lst:motivation} presents an example of an executor-related issue, where the \lstinline{HashJoin} operator is expected to run in parallel but fails to do so. We show another example of a cost model issue as follows.

\begin{figure}
\begin{lstlisting}[caption={Issue \blind{4} of cost model in PostgreSQL.},captionpos=t, label=lst:example2, escapeinside=&&, language=sql]
SELECT ... FROM CUSTOMER, WEB_SALES, DATE_DIM ...
WHERE ...t_w_secyear.sale_type = 'w'
   AND t_s_firstyear.dyear =  2001
   AND t_s_secyear.dyear = 2001+1
   AND t_c_firstyear.dyear =  2001
   AND t_c_secyear.dyear =  2001+1
   AND t_w_firstyear.dyear = 2001...;
------------------Code Manipulation------------------
--- a/src/backend/optimizer/util/pathnode.c
+++ b/src/backend/optimizer/util/pathnode.c
@@ -555,16 +555,6 @@ add_path(...)
  case COSTS_BETTER2:
-   if (keyscmp != PATHKEYS_BETTER1)
+   if (keyscmp == PATHKEYS_BETTER1)
    {
      ...
      accept_new = false; /* old dominates new */
    }
    break;
-------------Query Plans and Performance-------------
...                     ...
->NestLoop(cost=794037) ->HashJoin(cost=794032)
...                     ...
Time: 50m:24s:486ms     Time: 8s:179ms
\end{lstlisting}
\end{figure}

\paragraph{Example: an issue of cost model.}
\autoref{lst:example2} shows a performance issue \tool found in PostgreSQL with query 4 in the TPC-DS benchmark. This issue exposes the most significant performance gap (51 minutes to 8 seconds). Similarly, we show relevant clauses of the query in lines 1--7 and relevant code in lines 9-18. This query includes multi-table joining, so multiple possible query plans are available for various joining order and algorithms. In line 13, the \emph{IF} statement checks whether to exclude additional query plans for evaluation. For this query, the true branch in lines 15--18 is executed, and the chosen query plan uses \lstinline{NestLoop} to join tables. We manipulate the code to disable the true branch, and more potential query plans are evaluated. Therefore, a much more efficient query plan that uses \lstinline{HashJoin} to join tables is evaluated and chosen. 
Developers quickly confirmed this issue after we reported. 
After analysis, developers believe the root cause is the accuracy of selectivity estimation for the multiple predicates in the \lstinline{WHERE} clause. Fixing this performance issue is challenging because we lack the information on the joint column distribution.

\paragraph{False positive analysis.}
Two types of false positives exist in the performance issues found by \tool. We consider a reported issue a true positive only if it is confirmed by the developers; otherwise, it is treated as a false positive.
One source of false positives in \tool is misalignment with developers’ intended optimization trade-offs. A false positive \tool found in PostgreSQL involves the inline optimization for \emph{Common Table Expressions} (CTEs), which determines whether to execute the CTE and the outer query together or separately. Although \tool observed a 64$\times$ performance improvement when forcing inline optimization, PostgreSQL developers did not consider it a true issue. They explained that enabling inline improves performance for some workloads but degrades it for others. Since there is currently no algorithm to distinguish between these cases, the optimization remains off by default. This type of trade-off-based false positive occurred only twice, suggesting that it is rare for \tool.
Another potential source of false positives comes from performance fluctuation. We observed that DBMS performance can fluctuate due to environmental factors such as system load or cache state. In our initial experiments, such fluctuation incurred around 30\% false performance issues. To address this, we strictly controlled system usage and only reported issues when the observed performance gap exceeded 10\%. In this way, we did not observe any false positives caused by performance fluctuation, as confirmed by DBMS developers.

\result{\tool found \numbugs previously unknown performance issues on widely used benchmarks TPC-H and TPC-DS.}

\subsection*{Q.2 Necessity}

We evaluated whether \tool is necessary to find these performance issues. Specifically, we examined 1) whether the performance issues found by \tool can be potentially found by previous performance-testing methods, and 2) whether these performance issues existed when prior methods were proposed. We considered the same performance-testing methods in \autoref{sec:study}: \apollo~\cite{jung2019apollo}, \amoeba~\cite{liu2022automatic}, \cert~\cite{ba2024cert}, and \puppy~\cite{wu2024puppy}. 

\begin{table}
    \centering\footnotesize
    \caption{The reproducibility of previous methods on the issues found by \tool.}
    \begin{tabular}{@{}l@{}rcccc@{}}
        \toprule
        \textbf{DBMS} & \textbf{Issue ID} & \textbf{\apollo} & \textbf{\amoeba} & \textbf{\cert} & \textbf{\puppy} \\
        \midrule
PostgreSQL  & \blind{\href{https://www.postgresql.org/message-id/SEZPR06MB6494BD3DDF5B03700032C2C98A782\%40SEZPR06MB6494.apcprd06.prod.outlook.com}{1}  }  &        &        &        &        \\ 
PostgreSQL  & \blind{\href{https://www.postgresql.org/message-id/SEZPR06MB649480C104C2479D0F0D8ACD8A512\%40SEZPR06MB6494.apcprd06.prod.outlook.com}{2}  }  &        &        &        &        \\ 
PostgreSQL  & \blind{\href{https://www.postgresql.org/message-id/SEZPR06MB64943EFCD432E8069D610E458A4E2\%40SEZPR06MB6494.apcprd06.prod.outlook.com}{3}  }  & \cmark &        &        &        \\ 
PostgreSQL  & \blind{\href{https://www.postgresql.org/message-id/SEZPR06MB6494F6A2837995BDD4E0BF9A8A5F2\%40SEZPR06MB6494.apcprd06.prod.outlook.com}{4}  }  & \cmark &        &        &        \\ 
PostgreSQL  & \blind{\href{https://www.postgresql.org/message-id/SEZPR06MB649422CDEBEBBA3915154EE58A232\%40SEZPR06MB6494.apcprd06.prod.outlook.com}{5}   }  &        &        &        &        \\ 
MySQL       & \blind{\href{https://bugs.mysql.com/bug.php?id=116456}{116456}                                                                           }  &        &        &        &        \\ 
MySQL       & \blind{\href{https://bugs.mysql.com/bug.php?id=116484}{116484}                                                                           }  & \cmark &        &        &        \\ 
MySQL       & \blind{\href{https://bugs.mysql.com/bug.php?id=116309}{116309}                                                                           }  & \cmark &        &        &        \\ 
MySQL       & \blind{\href{https://bugs.mysql.com/bug.php?id=116534}{116534}                                                                           }  &        &        &        &        \\ 
MySQL       & \blind{\href{https://bugs.mysql.com/bug.php?id=116773}{116773}                                                                           }  &        &        &        &        \\ 
MySQL       & \blind{\href{https://bugs.mysql.com/bug.php?id=116774}{116774}                                                                           }  & \cmark &        &        &        \\ 
MySQL       & \blind{\href{https://bugs.mysql.com/bug.php?id=116775}{116775}                                                                           }  &        &        &        &        \\ 
MySQL       & \blind{\href{https://bugs.mysql.com/bug.php?id=116776}{116776}                                                                           }  & \cmark &        & \cmark &        \\ 
MySQL       & \blind{\href{https://bugs.mysql.com/bug.php?id=116777}{116777}                                                                           }  &        &        & \cmark &        \\ 
CockorachDB & \blind{\href{https://github.com/cockroachdb/cockroach/issues/134803}{134803}                                                             }  &        &        &        &        \\ 
CockorachDB & \blind{\href{https://github.com/cockroachdb/cockroach/issues/135001}{135001}                                                             }  &        &        & \cmark &        \\ 
CockorachDB & \blind{\href{https://github.com/cockroachdb/cockroach/issues/136350}{136350}                                                             }  & \cmark &        & \cmark &        \\ 
MariaDB     & \blind{\href{https://jira.mariadb.org/browse/MDEV-35280}{35280}                                                                          }  &        &        &        &        \\ 
MariaDB     & \blind{\href{https://jira.mariadb.org/browse/MDEV-35332}{35332}                                                                          }  &        &        &        &        \\ 
MariaDB     & \blind{\href{https://jira.mariadb.org/browse/MDEV-35331}{35331}                                                                          }  &        &        &        &        \\ 
MariaDB     & \blind{\href{https://jira.mariadb.org/browse/MDEV-35333}{35333} }  & \cmark &        &        & \cmark \\ 
        \bottomrule
        \multicolumn{2}{r}{\textbf{Sum (\cmark):}} & \textbf{8} & \textbf{0} & \textbf{4} & \textbf{1}
    \end{tabular}
    \label{tab:comparison}
\end{table}

\paragraph{Issue reproducing.}
We evaluated whether the performance issues found by \tool can be potentially found by prior performance-testing methods \apollo, \amoeba, \cert, and \puppy. Since these methods target different aspects of performance, such as cardinality estimation or configuration sensitivity, and \tool focuses on source code manipulation, the sets of bugs they uncover are not directly comparable. To address this, we conducted a qualitative analysis, a common approach in prior work~\cite{ba2024keep,Rigger2020TLP,Rigger2020NoREC}. Specifically, given a performance issue and both execution paths of a DBMS before and after \tool's code manipulation, 1) if the first lines of discrepancy execution paths are lastly updated in different commits, we assumed the issue can be found by \apollo; 2) if the instrumented \emph{IF} condition is directly controlled by a SQL clause, we assumed there exists a pair of semantic-equivalent queries to execute both paths so that the issue can be found by \amoeba; 3) if the first branches in discrepancy execution paths directly manipulate estimated cardinalities, we assumed the issue can be found by \cert; 4) if the instrumented \emph{IF} condition is directly controlled by an external configuration, we assumed the issue can be found by \puppy. These assumptions do not guarantee that prior methods can find these issues. For example, although the first lines of both execution paths are updated in different commits, it is not necessary that both versions of the DBMSs show the performance gap.

\paragraph{Results.}
\autoref{tab:comparison} shows whether prior methods can find the performance issues found by \tool. Overall, 8, 0, 4, and 1 issues have the potential to be found by \apollo, \amoeba, \cert, and \puppy, respectively. It demonstrates the necessity of \tool for finding these performance issues, and black-box methods are insufficient to explore optimization strategies for finding performance issues. The number of \apollo is much higher than others, because developers typically submit a minimal code block in each commit, making it more likely to observe the code update in different commits. 

\begin{figure*}
    \centering
    \includegraphics[width=\textwidth]{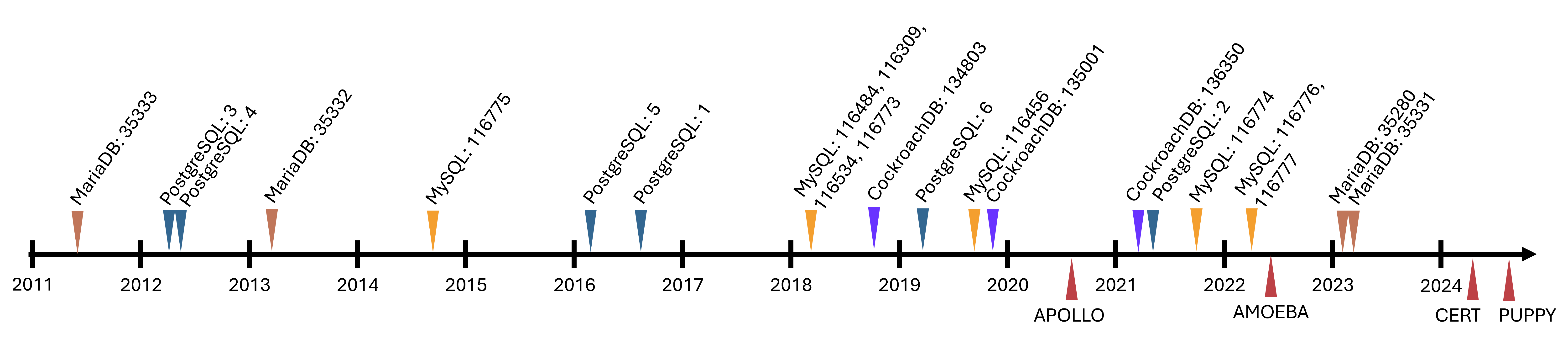}
    \caption{The last commit associated with the issues found by \tool and the publication date of prior methods.}
    \label{fig:timeline}
\end{figure*}

\paragraph{Issue existence date.}
Additionally, we evaluated whether these performance issues existed when prior performance-testing methods \apollo, \amoeba, \cert, and \puppy were proposed. If so, it demonstrates that these performance issues were not found by prior methods with an extensive testing effort. We retrieved the last commit date of the code manipulated \tool, and we assumed the performance issues existed no later than the date.

\paragraph{Results.}
\autoref{fig:timeline} shows the last commit date associated with the \numbugs performance issues found by \tool and the publication date of the prior methods. Issues \blind{\#116484}, \blind{\#116309}, \blind{\#116534}, and \blind{\#116773} share the same commit, as do issues \blind{\#116776} and \blind{\#116777}. It is because both commits focus on tidying up code style, resulting in extensive modifications, including the manipulation of code across multiple issues. In total, 14, 19, 21, and 21 of \numbugs issues existed before the publication date of \apollo, \amoeba, \cert, and \puppy. A large portion of performance issues already existed for a long time, while prior methods have not found them in practice.

\result{Most issues found by \tool cannot be found by \apollo, \amoeba, \cert, and \puppy, and existed before the publication of prior methods.}

\subsection*{Q.3 Sensitivity}

\paragraph{The soundness of code manipulations.}
\tool manipulates all \emph{IF} branches and ensures that the flipped branches impact only performance, not functionalities, through differential testing. To evaluate its soundness, we manually examined the branches that passed the functionality testing. Since the total number of branches (\eg, in MySQL and MariaDB) is too large for exhaustive analysis, we randomly sampled 100 flipped branches and reviewed their associated code and comments, which typically describe the purpose of each branch or function. Our manual inspection showed that none of the flipped branches affect the correctness of the results. For PostgreSQL, MySQL, CockroachDB, and MariaDB, 100\%, 92\%, 100\%, and 93\% of the sampled branches, respectively, control the optimization decisions, where modifications are expected to affect performance only. The remaining branches involved non-query operations such as table locking, file writing, and authentication. These operations do not influence query results and thus cannot be filtered out by differential testing. Because we instrumented all SQL-related code in MySQL and MariaDB, some non-optimization branches were included. However, as these non-query operations typically do not impact query results, they do not compromise the soundness of our code manipulation for finding performance issues.

\begin{table}
    \centering\small
    \caption{Average number of executed and all instrumented code manipulations by \tool.}
    \begin{tabular}{@{}lrrr@{}}
        \toprule
        \textbf{DBMS} & \textbf{TPC-H} & \textbf{TPC-DS} & \textbf{All} \\
        \midrule
         PostgreSQL & 1,202 (30.65\%)  & 1,362 (34.73\%) & 3,922 \\
         MySQL      & 2,285 (18.42\%)  & 2,404 (19.38\%) & 12,407 \\
         CockroachDB& 2,384 (32.09\%)  & 2,448 (32.95\%) & 7,429 \\
         MariaDB    & 3,833 (12.60\%)  & 3,947 (12.97\%) & 30,428 \\
        \bottomrule
        \textbf{Avg:} &  23.44\% & 25.01\% & \\
    \end{tabular}
    \label{tab:count}
\end{table}

\paragraph{The efficiency contribution of eliminating unnecessary code manipulations.}
For each query, \tool records the executed \emph{IF} statements and only manipulates them, reducing the workload associated with unnecessary code manipulations. We evaluated the extent to which unnecessary code manipulations can be excluded. \autoref{tab:count} presents the average number of executed and all instrumented code manipulations by \tool. On average, \tool executes only 23.44\% and 25.01\% of all instrumented code manipulations in the TPC-H and TPC-DS benchmarks, respectively. It significantly improves the testing efficiency by excluding the overhead of executing the remaining 76.56\% and 74.99\% unnecessary code manipulations.

\paragraph{The efficiency contribution of eliminating recompilation.}
\tool eliminates the need for recompilation by instrumenting the code to read environmental variables that control execution flow. To evaluate the impact of this instrumentation on testing efficiency, we specifically assessed whether the time required for recompilation represents a significant bottleneck. Our experiments show that even with full utilization of 112 CPU cores, compiling PostgreSQL, MySQL, CockroachDB, and MariaDB once takes 49.0, 655.0, 309.2, and 202.0 seconds, respectively. Given that our testing campaign involves a large number of executions, the overhead of recompiling for each execution becomes impractical. By leveraging \tool’s design, we eliminate recompilation overhead, thereby significantly enhancing testing efficiency.

\begin{figure}
    \centering
    \includegraphics[width=\columnwidth]{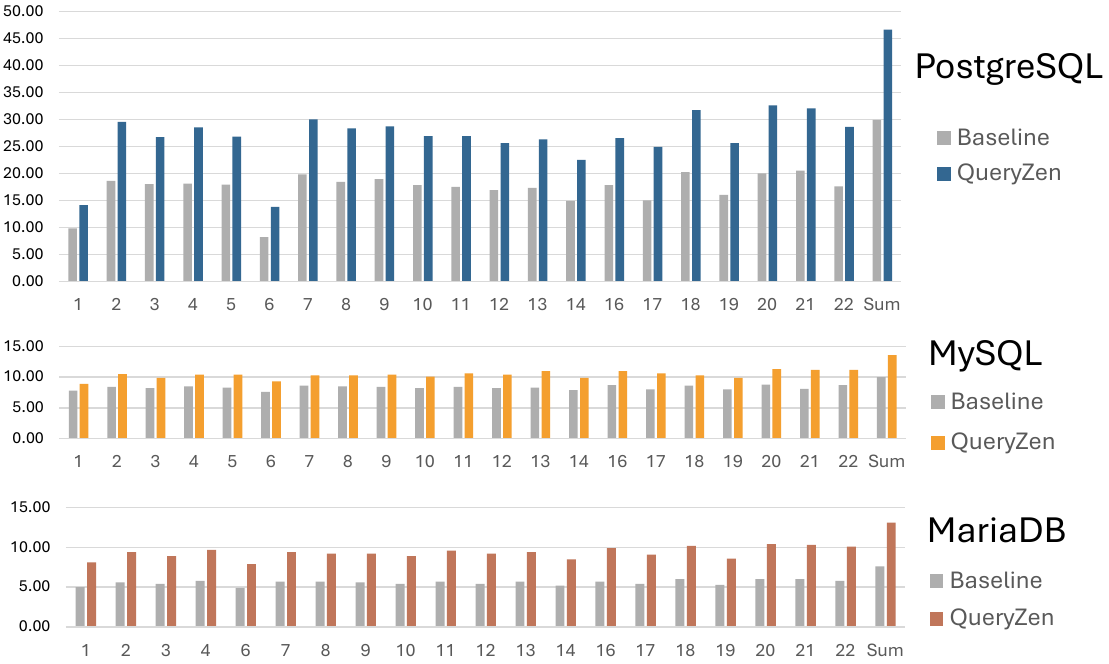}
    \caption{Branch coverage for TPC-H benchmark before and after manipulating code.}
    \label{fig:covtpch}
\end{figure}

\paragraph{Branch coverage increased by \tool.}
We evaluated the extent of optimization space explored. Specifically, we compared the branch coverage before and after code manipulation. A larger branch coverage gap indicates that \tool’s manipulations enable the exploration of more optimization opportunities. Our analysis focused on the query optimizer, which is the component we instrumented. We identified 35,522 branches in PostgreSQL, 531,247 branches in MySQL, and 564,585 branches in MariaDB query optimizers. The higher branch counts for MySQL and MariaDB are due to \tool instrumenting all SQL-related code. We did not evaluate CockroachDB, which does not support branch coverage measurement for end-to-end tests~\cite{cockroachdbcodecoverage}.

\autoref{fig:covtpch} shows the branch coverage for the TPC-H benchmark before and after manipulating code. In the x-axis, each number represents the query ID, and right-column \emph{Sum} represents the accumulated coverage of all queries. The number in the y-axis represents the percentage of branch coverage. 
\tool cumulatively covers 16.7\%, 3.6\%, and 5.5\% more branches than executing original queries in PostgreSQL, MySQL, and MariaDB. On average of all queries, \tool covers 3,339, 11,049, and 21,115 additional branches in the optimizers of PostgreSQL, MySQL, and MariaDB, accounting for 9.40\%, 2.08\%, and 3.74\% of all measured branches. The percentage is lower in MySQL and MariaDB because their total number of measured branches is significantly larger than PostgreSQL's.
We observed a similar result for the TPC-DS benchmark. \tool cumulatively covers 19.1\%, 4.6\%, and 8.1\% more branches than executing original queries in PostgreSQL, MySQL, and MariaDB. Notably, \tool covers 59.9\% of the PostgreSQL optimizer’s branches, demonstrating that a substantial portion of optimization strategies has been tested. On average of all queries, \tool covers 3,311, 8,818, and 22,922 additional branches in PostgreSQL, MySQL, and MariaDB, accounting for 9.32\%, 1.66\%, and 4.06\% of all measured branches.
Note that, during the branch coverage evaluation of MySQL, unexpected behaviors caused by manipulated code occasionally blocked data collection, resulting in incomplete results.
The results show \tool substantially explores optimization space. Using more diverse workloads and code manipulations may further expand coverage.

\section{Discussion}

\paragraph{Methodology.}
We propose a white-box testing methodology for identifying performance issues by directly manipulating program code. Unlike existing approaches~\cite{ba2024cert, liu2022automatic, wu2024puppy}, which focus on manipulating external inputs, \method operates on the program code itself. This approach offers several advantages. First, it enables systematic and fine-grained testing of program logic, overcoming the limitations of input-based methods that can only trigger behaviors in a constrained manner. As a result, our methodology uncovered \numbugs unique and previously unknown performance issues. Second, the code manipulated by \method facilitates issue debugging and analysis. Prior works compare the performance of executing input pairs, but it often remains unclear which part of the code incurs the performance difference. In contrast, our approach directly highlights the relevant code segments responsible for performance variations, streamlining the analysis process. Finally, \method does not require domain-specific knowledge. Traditional white-box testing methods, such as unit testing, often rely on domain expertise to establish appropriate testing entry points. In contrast, \method flips general code branches, making it readily applicable to target systems without requiring specialized knowledge.

\paragraph{Path to adoption.} 
We believe that a simple testing approach has the potential for wide adoption. From a conceptual perspective, \method is a general method to manipulate program code to find performance issues, which is easy to understand. From an implementation perspective, \method is easy to implement as we implemented the code manipulation script in around 200 lines of Python code and testing logic in around 300 lines of Shell script. Our implementation is not DBMS-specific, so we do not require additional implementation effort to test other DBMSs. From an integration perspective, \method can be paired with any workload. We chose TPC-H and TPC-DS benchmarks because they are representative of performance workloads. From an applicability perspective, \method can test any open-sourced DBMSs as \method is based on source code. Considering these features of \method, we argue that \method can be widely adopted. 

\paragraph{Fixing performance issues.} 
\method highlights the relevant code segment for the performance issue, facilitating analysis but fixing them still requires non-trivial effort. One reason is that fixing performance issues requires comprehensive consideration, which usually consumes much time. The code manipulated by \method shows performance issues based on a single workload without considering all workloads, so more effort is required to propose a fixing patch that does not lower the performance for other workloads. Another reason is that performance issues might have lower priority than other issues, such as correctness and crash bugs.

\paragraph{Limitations.}
While \method demonstrates effectiveness in identifying previously unknown performance issues, it has several limitations. First, although we apply functionality checks using differential testing and validate against thousands of test cases, semantic equivalence is inferred rather than guaranteed. Subtle functional divergences may go undetected without formal specifications, which would require manual effort from developers. Second, the effectiveness of \method depends on the quality of the underlying workload. While we use TPC-H and TPC-DS to represent common scenarios, other types of workloads might trigger different optimizations. Finally, our evaluation was conducted under a fixed system configuration. Although we observed consistent performance gaps across repeated trials, different hardware or memory settings may affect reproducibility.

\section{Related Work}


\paragraph{Testing performance issues.} 
The most related strand of research is on finding performance issues in DBMSs automatically. 
Ba \etal proposed \cert~\cite{ba2024cert} to find performance issues by testing specifically cardinality estimation. Jung \etal proposed \apollo~\cite{jung2019apollo}, which compares the execution times of a query on two versions of a database system to find performance regression bugs. Liu \etal proposed \amoeba~\cite{liu2022automatic}, which compares the execution time of a semantically-equivalent pair of queries to identify an unexpected slowdown. Wu \etal proposed \puppy~\cite{wu2024puppy}, which examines whether performance improves after adjusting default configurations. 
\add{In addition, several fuzzing-based approaches explore performance issues in general software. SlowFuzz~\cite{petsios2017slowfuzz} searches for slow executions by maximizing total execution path length, while PerfFuzz~\cite{lemieux2018perffuzz} uses multi-dimensional feedback to independently maximize execution counts of individual program locations, helping reveal code regions prone to excessive execution time.}
Unlike the above methods that examine optimization logic by executing different workloads and configurations, \method enforces the optimization logic by code manipulation for the same workload and configurations, achieving a fine-grained testing of optimization logic at the source code level.

\paragraph{Performance optimization.}
Performance optimization is a longstanding research problem in the system and database research community. Ruan \etal~\cite{ruan2023persist} uses disaggregated memory to improve DBMS performance. Bindschaedler \etal~\cite{bindschaedler2020hailstorm} disaggregated computation and storage in DBMSs to improve performance. Tsalapatis \etal proposed Memsnap~\cite{tsalapatis2024memsnap} to optimize the DBMS storage layer. Liu \etal~\cite{liu2024juno} proposed Juno to optimize the search algorithm in vector DBMSs. Apart from optimization, several benchmarks are used to compare performance across DBMSs. The \emph{TPC-H}~\cite{tpch} and \emph{TPC-DS}~\cite{tpcds} benchmarks are widely recognized as industry standards. Leis \etal introduced the \emph{Join Order Benchmark (JOB)}\cite{leis2015good}, which incorporates complex join orders.
Instead of proposing new algorithms and system designs to improve the performance, \method aims to expose the unexpected suboptimal performance due to implementation issues.

\add{
\paragraph{Configuration-based performance optimization.}
Systems typically expose configuration parameters that control execution, and searching for optimal configurations is a common approach to improving performance. Violet~\cite{hu2020automated} applies selective symbolic execution to detect specious configurations that unexpectedly degrade performance. LearnConf~\cite{li2020statically} investigates configuration-related performance bugs and develops a static analysis tool that identifies which configurations influence specific performance dimensions and in what ways. Jin \etal~\cite{jin2012understanding} conduct empirical studies and propose detection frameworks for performance bugs in real systems.
These works primarily focus on misconfigurations, whereas \method targets a broader class of performance issues in a more systematic way.
}


\paragraph{DBMS testing.} 
Beyond performance testing, automated methods have been developed to uncover various DBMS bugs. Many fuzzing tools target security-critical issues, such as memory errors. Tools like SQLSmith~\cite{sqlsmith}, Griffin~\cite{fu2022griffin}, DynSQL~\cite{jiangdynsql}, and ADUSA~\cite{liu2022automatic} employ grammar-based techniques to generate test cases aimed at detecting memory-related vulnerabilities. Inspired by grey-box fuzzers like AFL~\cite{afl}, Squirrel~\cite{zhong2020squirrel} uses code coverage feedback to identify memory issues.
Other approaches focus on discovering logic bugs in DBMSs. Oracles such as PQS~\cite{Rigger2020PQS}, NoREC~\cite{Rigger2020NoREC}, and TLP~\cite{Rigger2020TLP} detect logic bugs in \lstinline{SELECT} statement implementations, while DQE~\cite{song2023testing} targets logic errors in \lstinline{UPDATE} and \lstinline{INSERT} statements. 
Similar to testing approaches that find bugs, our core contribution is a novel testing methodology that finds performance issues by manipulating program code.

\section{Conclusion}
In this paper, we have proposed \emph{Branch Flip Analysis} (\method), a general \emph{white-box} method to find performance issues by flipping branches of \emph{IF} statements in source code. As the default optimization strategies should ideally have optimal performance, we disable or enforce an optimization by flipping code branches to derive a reference DBMS, whose performance is expected to be no better than the original DBMS. \method provides a new angle to find performance issues and can test target DBMSs in a systematic and fine-grained manner, facilitating issue analysis. Technically, we improved the testing efficiency by eliminating recompilation and unnecessary code manipulations. We implemented \method into the prototype \tool and demonstrated its effectiveness by evaluating it on four mature and widely used DBMSs, PostgreSQL, MySQL, CockroachDB, and MariaDB. \tool found \numbugs unique and previously unknown performance issues with standard benchmarks TPC-H and TPC-DS. These findings resulted in an average performance improvement of 26.2$\times$, and up to 374.9$\times$. We believe such a straightforward methodology helps improve DMBSs’ performance in practice and opens a promising research avenue to analyze program behaviors by manipulating source code, instead of manipulating external inputs so that target systems can be systematically examined.

\bibliographystyle{plain}
\bibliography{references}

\end{document}